\documentclass[twocolumn,amsmath,amssymb]{revtex4}
\usepackage{bm}
\usepackage{graphicx}
\usepackage{epstopdf}
\usepackage{amsmath}

\begin{document}

\newcommand{\uu}[1]{\underline{#1}}
\newcommand{\pp}[1]{\phantom{#1}}
\newcommand{\be}{\begin{eqnarray}}
\newcommand{\ee}{\end{eqnarray}}
\newcommand{\ve}{\varepsilon}
\newcommand{\vs}{\varsigma}
\newcommand{\Tr}{{\,\rm Tr}}
\newcommand{\pol}{\frac{1}{2}}
\newcommand{\RR}{\rotatebox[origin=c]{180}{$\mathbb{R}$} }
\newcommand{\CC}{\rotatebox[origin=c]{180}{$\mathbb{C}$} }
\newcommand{\rr}{\mathbb{R}}
\newcommand{\Exp}{{\,\rm Exp\,}}
\newcommand{\Sin}{{\,\rm Sin\,}}
\newcommand{\Cos}{{\,\rm Cos\,}}
\newcommand{\Tan}{{\,\rm Tan\,}}
\newcommand{\Sinh}{{\,\rm Sinh\,}}
\newcommand{\Cosh}{{\,\rm Cosh\,}}

\title{
Non-Newtonian mathematics instead of non-Newtonian physics: Dark matter and dark energy from a mismatch of arithmetics}
\author{Marek Czachor}
\affiliation{
Katedra Fizyki Teoretycznej i Informatyki Kwantowej,
Politechnika Gda\'nska, 80-233 Gda\'nsk, Poland
}
\begin{abstract}
Newtonian physics is based on Newtonian calculus applied to Newtonian dynamics. New paradigms such as `modified Newtonian dynamics' (MOND)  change the dynamics,  but do not alter the calculus. However, calculus is dependent on arithmetic, that is the ways we add and multiply numbers. For example, in special relativity we add and subtract velocities by means of addition $\beta_1\oplus \beta_2=\tanh\big(\tanh^{-1}(\beta_1)+\tanh^{-1}(\beta_2)\big)$,  although multiplication $\beta_1\odot \beta_2=\tanh\big(\tanh^{-1}(\beta_1)\cdot\tanh^{-1}(\beta_2)\big)$, and division $\beta_1\oslash \beta_2=\tanh\big(\tanh^{-1}(\beta_1)/\tanh^{-1}(\beta_2)\big)$ do not seem to appear in the literature. The map $f_\mathbb{X}(\beta)=\tanh^{-1}(\beta)$  defines an isomorphism of the arithmetic in $\mathbb{X}=(-1,1)$ with the standard one in $\mathbb{R}$. The new arithmetic is projective and non-Diophantine in the sense of Burgin (1977), while ultrarelativistic velocities are super-large in the sense of Kolmogorov (1961). Velocity of light plays a role of non-Diophantine infinity. The new arithmetic allows us to define the corresponding derivative and integral, and thus a new calculus which is non-Newtonian in the sense of Grossman and Katz (1972). Treating the above example as a paradigm, we ask what can be said about the set $\mathbb{X}$  of `real numbers', and the isomorphism $f_{\mathbb{X}}:\mathbb{X}\to \mathbb{R}$,  if we assume the standard form of Newtonian mechanics and general relativity (formulated by means of the new calculus) but demand agreement with astrophysical observations. It turns out that the observable accelerated expansion of the Universe 
can be reconstructed with zero cosmological constant if $f_\mathbb{X}(t/t_H)\approx 0.8\sinh (t-t_1)/(0.8\, t_H)$. The resulting non-Newtonian model is exactly quivalent to the standard Newtonian one with $\Omega_\Lambda=0.7$, $\Omega_M=0.3$. Asymptotically flat rotation curves are obtained if `zero', the neutral element $0_{\mathbb{X}}$ of addition,  is nonzero from the point of view of the standard arithmetic of $\mathbb{R}$. This  implies $f^{-1}_{\mathbb{X}}(0)=0_{\mathbb{X}}>0$. 
The opposition Diophantine vs. non-Diophantine, or Newtonian vs. non-Newtonian, is an arithmetic analogue of Euclidean vs. non-Euclidean in geometry. We do not yet know if the proposed generalization ultimately removes any need of  dark matter, but it will certainly change estimates of its parameters. Physics of the dark universe seems to be both geometry and arithmetic.

\end{abstract}
\maketitle

\section{Dark Universe and its dark arithmetic}

\begin{flushright}
{\it The difficulty lies more in the notions themselves 

than in the construction\/}

Bernhard Riemann (1854)

\end{flushright}

\medskip

Two hundred years ago the very idea of abandoning Euclidean axioms of geometry seemed so self-contradictory that even C.~F.~Gauss, a mathematician of highest reputation, found it imprudent to publish his thoughts on the subject. Euclid's fifth axiom of parallels was a truism for contemporaries of Bolyai, Lobachevski and Gauss, and we basically understand why: their everyday experiences were small-scale.  Nowadays, not only are we all accustomed to non-Euclidean geometries, but we even find it difficult to think of gravitational physics in categories different from just a geometry. 

Yet, modern space-time physics is clearly at crossroads. The experimental value of the cosmological constant is some $10^{120}$ smaller than its theoretical estimate \cite{Weinberg} ---  probably the worst disagreement between theory and experiment in history of science. A radical change of paradigm should not be a surprise. 

The goal of this paper is to draw the attention of the dark-universe community to an overlooked mathematical freedom: the axioms of arithmetic. Problems with dark energy and dark matter may indicate that physics is geometry... and arithmetic.

To begin with, many would probably agree that if we were to give an example of an absolute and self-evident truth, one would mention $2+2=4$. Now, is it as obvious as the axiom of parallels, or perhaps `more obvious'? Is 
$$
2^{100^{100^{100}}}+2^{100^{100^{100}}}=2^{100^{100^{100}}+1}
$$ 
equally obvious? Has anybody any practical experience with adding numbers that big? Even supercomputers cannot process numbers greater than the so-called machine infinity, a finite number $N_\infty$ which does not increase if we add 1 to it. So, $N_\infty<\infty$ and $N_\infty+1=N_\infty$. This type of arithmetic is either inconsistent, or {\it non-Diophantine\/}. The later means that some of the rules of arithmetic, formalized by Diophantos of Alexandria, may have to be dropped. Similarly to the rules of geometry, formalized by Euclid of Alexandria.

To put what I write in a wider context let me mention that A.~N.~Kolmogorov himself proposed to split natural numbers into classes of small, medium, large, and super-large, and each class might in principle be based on different rules, dependent on our computing capabilities. Kolmogorov expressed his views in two papers addressed to high-school pupils \cite{Kolmogorov}. We do not know if he had any concrete mathematical system in mind. A step further was done by P.~K.~Rashevsky \cite{Rashevsky} whose letter to Uspekhi Matematicheskich Nauk explicitly formulated the program of going beyond the `dogma' of natural numbers. Similarly to Kolmogorov, Rashevsky did not propose any concrete non-Diophantine system of axioms. The first explicitly non-Diophantine arithmetic of natural numbers was described in the same journal four years later by Mark S. Burgin \cite{Burgin77,Burgin,Burgin2,Burgin3}. 

Independently of the efforts of Rashevsky and Burgin, M.~Grossman and R.~Katz worked out a form of calculus which culminated in their little book {\it  Non-Newtonian Calculus\/} \cite{GK,G79,G83}. It went basically unnoticed by the mainstream mathematical community, and was completely ignored by physicists. Their main idea was rediscovered two decades later by E.~Pap in his g-calculus \cite{Pap1993,Pap2008,G}. Another two decades later, but in its currently most general form, it was  rediscovered by myself \cite{MC2015,ACK2016a,ACK2016b,CzachorDE,ACK2018,Czachor2019}.

The term `non-Newtonian' refers here to the level of {\it calculus\/}, and not to the {\it laws of physics\/}. 
It should not be confused with Milgrom's `modified Newtonian dynamics'  (MOND) \cite{MOND,MOND2,B}, Moffat's `modified gravity' (MOG) \cite{MOG,MOG2}, or similar theories. A non-Newtonian calculus is based on a non-Diophantine arithmetic. The Newtonian calculus is based on the Diophantine arithmetic.  But Newton's equation relating force and acceleration, the three Newton laws of dynamics, or the `inverse square' Newton law of gravity remain unchanged. 
Theoretical freedom is hidden in various possible meanings of `plus', `times', `minus', `divided by', `squared'... This is the new paradigm.

Non-Diophantine arithmetic and the non-Newtonian calculus it implies automatically lead to two types of `dark universes': the ones where super-small and super-large physical quantities behave differently even though they satisfy the same physical laws, and those  identified with zero-measure sets whose physics is equipped with the usual laws, but which are invisible from the point of view of standard quantum measurements. 

In the present paper we will discuss simple examples illustrating each of the above concepts. In Sec.~\ref{Sec 2} we begin with a concrete arithmetic of a Burgin type which naturally splits real numbers into small, large and super-large. We note that relativistic velocities are in this sense super-large. Then in Sec.~\ref{Sec 3} we briefly explain the idea of non-Newtonian differentiation and integration. In the next section we combine the ideas from Sec.~\ref{Sec 2} and Sec.~\ref{Sec 3} and show that the standard Friedman equation without dark energy in fact {\it can\/} imply an accelerated expansion of the Universe. In Sec.~\ref{Sec 5} and Sec.~\ref{App} we show how to derive a `dark-matter' type of asymptotically flat velocity curve by means of the standard Newton equation of motion for a `$1/r$' potential. The non-Newtonian general prediction is briefly compared with the MOND paradigm in Sec.~\ref{Sec MOND}. In all the above examples the trick lies in a mismatch between the arithmetic employed in our modeling, and the arithmetic employed by the Universe. The problems with dark energy and dark matter look like an experimental indication that the arithmetic we all work with is not necessarily the one preferred by Nature. This is the main message of the paper. 

In Sec.~\ref{strange} we return to the dilemmas of the 19th century thinkers. We stress that we {\it do \/} perceive non-Diophantine arithmetic in our everyday life, but typically being unaware of it. Finally, in the Appendix we show how to formulate the issue of dark energy as an eigenvalue problem for a quantum system that `lives' in a set of zero Lebesgue measure, namely in a Cantor-dust fractal.

\section{Non-Diophantine arithmetic:
An example}
\label{Sec 2}

Let the set $X$ of physical variables have some physical dimension (length, say) and let $\ell$ be a fundamental unit. Consider the set $\mathbb{X}$ of dimensionless numbers obtained by dividing elements of $X$ by $\ell$, that is $\mathbb{X}=\{x=a/\ell, a\in X\}$. We assume that $\mathbb{X}$ has the same cardinality as the continuum $\mathbb{R}$. Just to have a feel of the generality we have at our disposal think of the following examples: $\mathbb{R}$ itself, the open unit interval 
$(0,1)$, the three-dimensional space $\mathbb{R}^3$, the Minkowski space, a Cantor dust, a Sierpi\'nski triangle, a Koch curve... All these sets have the same cardinality as reals, and therefore there exist one-to-one maps $f$ mapping them onto $\mathbb{R}$. Typically these maps are quite bizarre and discontinuous in metric topologies of $\mathbb{X}$ (try to invent a one-to-one $f:\mathbb{R}^3\to \mathbb{R}$), but this is not a problem, even if calculus is concerned.

Now, consider a bijection  $f: \mathbb{X}\to \mathbb{R}$ and the arithmetic in $\mathbb{X}$ induced by $f$,
\be
x\oplus x' &=& f^{-1}\big(f(x)+f(x')\big),\label{ar1}\\
x\ominus x' &=& f^{-1}\big(f(x)-f(x')\big),\label{ar2}\\
x\odot x' &=& f^{-1}\big(f(x)\cdot f(x')\big),\label{ar3}\\
x\oslash x' &=& f^{-1}\big(f(x)/f(x')\big).\label{ar4}
\ee
The bijection is a field isomorphism of $\mathbb{X}$ and $\mathbb{R}$,
\be
f(x\oplus x') &=& f(x)+f(x'),\\
f(x\odot x') &=& f(x)\cdot f(x').
\ee
For this reason $\oplus$ and $\odot$ are associative and commutative, and $\odot$ is distributive with respect to $\oplus$ \cite{Ex1}. 
Any such $\mathbb{X}$ is also ordered: $x\leq_\mathbb{X} x'$ if and only if $f(x)\leq f(x')$. 
The neutral elements of addition and multiplication read, respectively, $0_\mathbb{X}=f^{-1}(0)$ and  $1_\mathbb{X}=f^{-1}(1)$. Indeed, for any $x\in \mathbb{X}$
\be
x\oplus 0_\mathbb{X} = f^{-1}\big(f(x)+f(0_\mathbb{X})\big)=f^{-1}\big(f(x)+0\big)=x,\\
x\odot 1_\mathbb{X} = f^{-1}\big(f(x)\cdot f(1_\mathbb{X})\big)=f^{-1}\big(f(x)\cdot 1\big)=x.
\ee
$0_\mathbb{X}$ and $1_\mathbb{X}$ can be generalized to arbitrary natural numbers $n\in\mathbb{N}$, 
\be
n_\mathbb{X}=
\underbrace{1_\mathbb{X}\oplus \dots \oplus 1_\mathbb{X}}_{\textrm{$n$ times}}=f^{-1}(n),
\ee
and to any real numbers, since for  $r,s\in\mathbb{R}$
\be
r_\mathbb{X}\oplus s_\mathbb{X}=(r+s)_\mathbb{X}
\ee
if one defines $r_\mathbb{X}=f^{-1}(r)$, $s_\mathbb{X}=f^{-1}(s)$. 
So, $2_\mathbb{X}\oplus 2_\mathbb{X}=4_\mathbb{X}$, but typically $2_\mathbb{X}\neq 2$ and $4_\mathbb{X}\neq 4$. To put it differently, even if $2_\mathbb{X}= 2$ it does not yet mean that $4_\mathbb{X}= 4$.

To focus our attention let us imagine that $X$ is the open interval $(-L/2,L/2)$, $\mathbb{X}=\big(-L/(2\ell),L/(2\ell)\big)$, 
\be
f(x) &=& \frac{L}{\pi \ell}\tan\frac{\pi \ell x}{L}\quad \textrm{($\approx x$ for small $x$)},\\
f^{-1}(r) &=& \frac{L}{\pi\ell}\arctan\frac{\pi \ell r}{L} \quad \textrm{($\approx r$ for small $r$)}.
\ee
The neutral elements are 
\be
0_\mathbb{X} &=& f^{-1}(0)=0,\\
1_\mathbb{X} &=& f^{-1}(1)= \frac{L}{\pi\ell}\arctan\frac{\pi \ell}{L}
\ee
Taking $L$ at the order of the radius of the visible Universe, $L=8\times 10^{26}$~m, and $\ell$ at the order of the Planck length, $\ell=2\times 10^{-35}$~m, we find $1_\mathbb{X}\approx 1$. For bigger numbers we find
\be
n_\mathbb{X} &=& f^{-1}(n)= \frac{L}{\pi\ell}\arctan\frac{\pi \ell n}{L}\\
&=&
\frac{4\times 10^{61}}{\pi}\arctan\frac{\pi  n}{4\times 10^{61}}.
\ee
Notice that the available real numbers are limited by
\be
|r_\mathbb{X}|< 2\times 10^{61}=L/(2\ell).
\ee
The upper bound $L/(2\ell)$ plays the same role as the machine infinity $N_\infty$. The arithmetic we have constructed is consistent but non-Diophantine. 
Numbers such as $10^{55}$ are not small, but still medium-large in the sense of Kolmogorov, since
\be
10^{55}\oplus 10^{55}=
f^{-1}\big(f(10^{55})+ f(10^{55})\big)\approx 2.\times 10^{55},\label{1+1}
\ee
\be
10^{55}\oplus 10^{55}\oplus 10^{55}
&=&
f^{-1}\big(f(10^{55})+f(10^{55})+ f(10^{55})\big)
\nonumber\\
&\approx& 3.\times 10^{55}.\label{1+1+1}
\ee
The symbol of approximate equality $\approx$ in (\ref{1+1})-(\ref{1+1+1}) is here practically determined by the computing capabilities  of Wolfram Mathematica. 
However $10^{61}$ is already Kolmogorovian super-large:
\be
10^{61}\oplus 10^{61}
&=&
f^{-1}\big(f(10^{61})+ f(10^{61})\big)
\nonumber\\
&\approx& 1.40967\times 10^{61},
\ee
\be
10^{61}\oplus 10^{61}\oplus 10^{61}
&=&
f^{-1}\big(f(10^{61})+f(10^{61})+ f(10^{61})\big)
\nonumber\\
&\approx& 1.59033\times 10^{61}.
\ee
An arithmetic defined by a bijection $f$ is, in the terminology of Burgin, an example of a projective arithmetic with projection $f$ and coprojection $f^{-1}$. 

It is evident that relativistic addition of dimensionless velocities $\beta=v/c$ is also an example of projective-arithmetic non-Diophantine addition:  $X=(-c,c)$, $\mathbb{X}=(-1,1)$, $f(x)=\tanh^{-1}(x)$, $f^{-1}(x)=\tanh x$. The neutral elements are $0_\mathbb{X}=f^{-1}(0)=\tanh 0=0$, $1_\mathbb{X}=f^{-1}(1)=\tanh 1=(e^2-1)/(e^2+1)\approx 0.76$. Does $v=0.76 c$ have any special physical meaning? Non-Diophantine infinity equals $\infty _\mathbb{X}=f^{-1}(\infty)=1$.
Velocity of light is infinite, at least in the non-Diophantine sense.

\section{Non-Newtonian differentiation and integration}
\label{Sec 3}

Consider two sets $\mathbb{X}$, $\mathbb{Y}$, with arithmetics 
$\{\oplus_\mathbb{X},\ominus_\mathbb{X},\odot_\mathbb{X},\oslash_\mathbb{X},\le_\mathbb{X}\}$ and 
$\{\oplus_\mathbb{Y},\ominus_\mathbb{Y},\odot_\mathbb{Y},\oslash_\mathbb{Y},\le_\mathbb{Y}\}$,
respectively. A function  $A:\mathbb{X}\to \mathbb{Y}$ defines a new function $\tilde A:\mathbb{R}\to \mathbb{R}$ such that the diagram 
\be
\begin{array}{rcl}
\mathbb{X}                & \stackrel{A}{\longrightarrow}       & \mathbb{Y}               \\
f_\mathbb{X}{\Big\downarrow}   &                                     & {\Big\downarrow}f_\mathbb{Y}   \\
\mathbb{R}                & \stackrel{\tilde A}{\longrightarrow}   & \mathbb{R}
\end{array}\label{diagram}
\ee
is commutative. The neutral elements of addition read $0_\mathbb{X}=f^{-1}_\mathbb{X}(0)$, 
$0_\mathbb{Y}=f^{-1}_\mathbb{Y}(0)$, where we assume continuity 
$\lim_{r\to 0_-}f^{-1}_\mathbb{X}(r)=\lim_{r\to 0_+}f^{-1}_\mathbb{X}(r)$ and 
$\lim_{r\to 0_-}f^{-1}_\mathbb{Y}(r)=\lim_{r\to 0_+}f^{-1}_\mathbb{Y}(r)$ of the inverse bijections, so that $0_\mathbb{X}$ and $0_\mathbb{Y}$ are unambiguously defined.

The derivative of $A$ is defined as
\be
\frac{{\rm D}A(x)}{{\rm D}x}
&=&
\lim_{h\to 0}\Big(
A(x\oplus_\mathbb{X} h_\mathbb{X})\ominus_\mathbb{Y}A(x)\Big)\oslash_\mathbb{Y} h_\mathbb{Y},
\label{DA}
\ee
where the limit is appropriately constructed \cite{ACK2018,Czachor2019}. Notice that this is just the standard undergraduate-course definition of a derivative, but formulated in terms of a general addition, subtraction, and division. One proves that (\ref{DA}) implies
\be
\frac{{\rm D}A(x)}{{\rm D}x}
&=&
f_\mathbb{Y}^{-1}
\left(
\frac{{\rm d}\tilde A\big(f_\mathbb{X}(x)\big)}{{\rm d}f_\mathbb{X}(x)}
\right).
\label{DA'}
\ee
Here ${\rm d}\tilde A(r)/{\rm d}r$ is the usual Newtonian derivative of $\tilde A:\mathbb{R}\to \mathbb{R}$, and we assume of course that the latter is differentiable. The form (\ref{DA'}) is an alternative form of the definition of the non-Newtonian derivative, which is extremely useful in practical calculations. Readers interested in explicit examples involving different choices of arithmetics in $\mathbb{X}$ or $\mathbb{Y}$ should look into \cite{Czachor2019}.

The non-Newtonian derivative is linear with respect to $\oplus_\mathbb{Y}$ and satisfies the Leibniz rule  
\be
&{}&
\frac{{\rm D}\big(A_1(x)\odot_\mathbb{Y} A_2(x)\big)}{{\rm D}x}
=
\left(
A_1(x)\odot_\mathbb{Y} 
\frac{{\rm D}A_2(x)}{{\rm D}x}
\right)\oplus_\mathbb{Y}
\nonumber\\
&{}&
\pp{\frac{{\rm D}\big(A_1(x)\odot_\mathbb{Y} A_2(x)\big)}{{\rm D}x}=}
\left(
\frac{{\rm D}A_1(x)}{{\rm D}x}\odot_\mathbb{Y} A_2(x)
\right).
\ee
One also proves a chain rule for compositions of functions \cite{Czachor2019} which, in particular, implies
\be
\frac{{\textrm D}f_{\mathbb{X}}(x)}{{\textrm D}x}
= 1=
\frac{{\textrm D}f_{\mathbb{Y}}(x)}{{\textrm D}x},
\ee
\be
\frac{{\textrm D}f_{\mathbb{X}}^{-1}(x)}{{\textrm D}x}
= 1_{\mathbb{X}},\quad
\frac{{\textrm D}f_{\mathbb{Y}}^{-1}(x)}{{\textrm D}x}
=
1_{\mathbb{Y}}.
\ee

The bijections themselves are therefore always differentiable with respect to the derivatives they define, while the resulting derivatives are always equal to appropriate unit elements. This is true also in cases where the domains $\mathbb{X}$ and the images $A(\mathbb{X})$ are highly nontrivial sets such as fractals. Although typically such bijections are discontinuous in metric topologies of $\mathbb{X}$ and $\mathbb{Y}$,  it is enough that they are always continuous in topologies they induce in $\mathbb{X}$ and $\mathbb{Y}$ from the open-interval topology of 
$\mathbb{R}$.iedn

Once we have the derivatives we define a non-Newtonian (Riemann, Lebesgue,...) integral of  $A$ by
\be
\int_a^b A(x){\textrm D}x
&=&
f_\mathbb{Y}^{-1}
\left(
\int_{f_\mathbb{X}(a)}^{f_\mathbb{X}(b)}\tilde A(r){\textrm d}r
\right),\label{integr}
\ee
i.e. in terms of the Newtonian (Riemann, Lebesgue,...) integral of $\tilde A$. The two functions $A$ and $\tilde A$ are related by (\ref{diagram}). Under standard assumptions about differentiability and continuity of $\tilde A$ we obtain both fundamental theorems of non-Newtonian calculus, relating derivatives and integrals. 

Let us note that (\ref{integr}) reduces an integral over $\mathbb{X}$ to a 1-dimensional integral over $[f_\mathbb{X}(a),f_\mathbb{X}(b)]\subset \mathbb{R}$. So, for example, if $\mathbb{X}=\mathbb{R}^3$ we reduce a 3-dimensional integral to a 1-dimensional one. The clue that such a counterintuitive possibility exists can be found already in Wiener's lectures on Fourier analysis  \cite{Wiener}.

Any model which is usually formulated in terms of the Diophantine arithmetic and the Newtonian calculus can be regarded as a special case, with $f_\mathbb{X}(x)=x$ and $f_\mathbb{Y}(y)=y$, of a general projective-arithmetic non-Diophantine and non-Newtonian one. All physical theories have their non-Newtonian generalizations. 

\section{Arithmetic analogue of dark energy}
\label{Sec 4}

If the arithmetic we employ in mathematical modeling of physical theories is identical to some putative Objective Arithmetic of the Universe, then there is no possibility of verifying if the physical arithmetic is Diophantine or not. Simply, we will always find 
`two plus two equals four' and the like. We will not know that `two', `four' or `plus' implicitly involve some $f_\mathbb{X}$, so should be written with some subscript $\mathbb{X}$. However, what if the Universe `works' with some other arithmetic, not necessarily the one we are accustomed to? In principle, we can discover a mismatch between the two arithmetics, and thus discover a nontrivial $f_\mathbb{X}$ \cite{CzachorDE}.

\subsection{Matter dominated universe}

\begin{figure}
\includegraphics[width=8 cm]{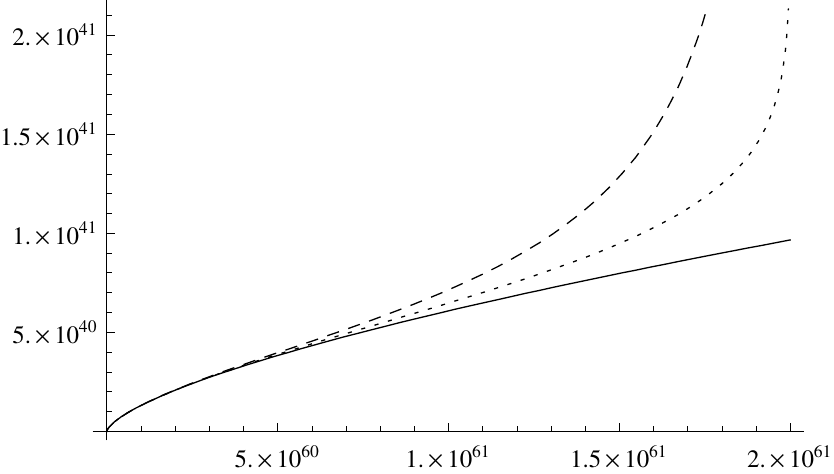}
\caption{Plots  of $a(t)$ (Eq. (\ref{a(t)}), full) and $A(t)$  (Eq. (\ref{A(t)})) for $\omega=1$. The dashed and dotted curves correspond to arithmetics given by (\ref{aritm1}) and (\ref{aritm2}), respectively. The models involve no dark energy.}
\label{fig1}
\end{figure}

Let us illustrate the phenomenon by  the Friedman equation
\be
\frac{{\rm d}a(t)}{{\rm d}t}
=
\omega/ a(t)^{1/2},
\ee
for a flat, matter dominated FRW model with exactly vanishing cosmological constant \cite{Hartle}. With the initial condition $a(0)=0$ we get 
\be
a(t)=(3\omega t/2)^{2/3}.\label{a(t)}
\ee
The non-Newtonian generalization reads for $A:\mathbb{X}\to \mathbb{Y}$,
\be
\frac{{\rm D}A(t)}{{\rm D}t}
=
\omega_\mathbb{Y} \oslash_\mathbb{Y} A(t)^{(1/2)_\mathbb{Y}},\quad
A(0_\mathbb{X})=0_\mathbb{Y},\label{F}
\ee
where $A^{(1/2)_\mathbb{Y}}\otimes_\mathbb{Y} A^{(1/2)_\mathbb{Y}}=A$, i.e.
\be
A^{(1/2)_\mathbb{Y}}=f^{-1}_\mathbb{Y}\left(\sqrt{f_\mathbb{Y}(A)}\right).
\ee
Employing the diagram (\ref{diagram}) we rewrite (\ref{F}) as 
\be
f_\mathbb{Y}^{-1}
\left(
\frac{{\rm d}\tilde A\big(f_\mathbb{X}(t)\big)}{{\rm d}f_\mathbb{X}(t)}
\right)
&=&
f_\mathbb{Y}^{-1}
\left(
\frac{f_\mathbb{Y}(\omega_\mathbb{Y})}{\tilde A\big(f_\mathbb{X}(t)\big)^{1/2}}
\right),
\ee
so that 
\be
\tilde A\big(f_\mathbb{X}(t)\big)
&=&
\big(3f_\mathbb{Y}(\omega_\mathbb{Y})f_\mathbb{X}(t)/2\big)^{2/3},\\
A(t)
&=&
f_\mathbb{Y}^{-1}
\Big(
\big(3f_\mathbb{Y}(\omega_\mathbb{Y})f_\mathbb{X}(t)/2\big)^{2/3}
\Big).\label{A(t)}
\ee
Now  let $\mathbb{X}=\mathbb{Y}=\big(-L/(2\ell),L/(2\ell)\big)$, $L=8\times 10^{26}$~m, $\ell=2\times 10^{-35}$~m,
$\omega_\mathbb{Y}=f_\mathbb{Y}^{-1}(\omega)$. Let the arithmetic be given by the example from Sec.~\ref{Sec 2},
\be
f_\mathbb{X}(x) = f_\mathbb{Y}(x)= \frac{L}{\pi \ell}\tan\frac{\pi \ell x}{L}.\label{aritm1}
\ee
The resulting $A(t)$ is shown in Fig.~\ref{fig1} (the dashed curve). For super-large times $A(t)$ bends up in a characteristic way, typical of dark-energy models of accelerating Universe. It is instructive to discuss also the $\tanh^{-1}$ we mentioned in the abstract. Therefore, let
\be
f_\mathbb{X}(x) = f_\mathbb{Y}(x)= \frac{L}{2\ell}\tanh^{-1}\frac{2\ell x}{L}.\label{aritm2}
\ee
The resulting $A(t)$ is shown in Fig.~\ref{fig1} (the dotted curve). In both cases the `dark energy' effect is of purely arithmetic origin. However, the `dark energy' behavior of $a(t)$ is here only qualitatively similar to the exact $\Omega_M=0.3$, $\Omega_\Lambda=0.7$ data \cite{DE1,DE2}. 

So, is there a  kind of arithmetic which is exactly compatible with the data? The answer is in the next subsection.

\subsection{Arithmetic behind $\Omega_M=0.3$, $\Omega_\Lambda=0.7$} 

The case $\Omega_M=0.3$, $\Omega_\Lambda=0.7$ involves  the Friedman equation,
\be
\frac{{\rm d}a(t)}{{\rm d}t}
=
\sqrt{\Omega_\Lambda a(t)^2 +\frac{\Omega_M}{ a(t)}},\quad a(t)>0,\label{FR1}
\ee
for a dimensionless scale factor evolving in a dimensionless time. Eq. (\ref{FR1}) is solved by
\be
a(t)
=
\left(
\sqrt{\frac{\Omega_M}{\Omega_\Lambda}}\sinh \frac{3\sqrt{\Omega_\Lambda}(t-t_1)}{2}
\right)^{2/3},\quad t> t_1.\label{a(t)1}
\ee
The dimensionless time is here expressed in units of the Hubble time $t_H\approx 13.58\times 10^9$~yr. 
The present time, $t=t_0$, satisfies $a(t_0)=1$ and thus
\be
t_0-t_1= \frac{2}{3\sqrt{\Omega_\Lambda}}\sinh^{-1}\sqrt{\frac{\Omega_\Lambda}{\Omega_M}}\approx 0.96.
\ee

For $\Omega_\Lambda\approx 0$, $t_1=0$,  we reconstruct (\ref{a(t)}) with $\omega=\sqrt{\Omega_M}$,
\be
a(t)
\approx
\left(
3\sqrt{\Omega_M}t/2
\right)^{2/3}.
\ee
A comparison of (\ref{a(t)1}), written as
\be
a(t)
&=&
\left[
\frac{3}{2}\sqrt{\Omega_M}\frac{2}{3\sqrt{\Omega_\Lambda}}\sinh \frac{3\sqrt{\Omega_\Lambda}}{2}(t-t_1)
\right]^{2/3},
\ee
with (\ref{A(t)}),
\be
A(t)
&=&
f_\mathbb{Y}^{-1}
\left[
\left(\frac{3}{2}f_\mathbb{Y}(\omega_\mathbb{Y})f_\mathbb{X}(t)\right)^{2/3}
\right],\label{A(t)1}
\ee
suggests $f_\mathbb{Y}(y)=y$, and 
\be
f_\mathbb{X}(t)
&=&
\frac{2}{3\sqrt{\Omega_\Lambda}}\sinh \frac{3\sqrt{\Omega_\Lambda}}{2}(t-t_1)\label{43}\\
&\approx& 
0.8\, \sinh \frac{t-t_1}{0.8},\\
f^{-1}_\mathbb{X}(r)
&=&
t_1+\frac{2}{3\sqrt{\Omega_\Lambda}}\sinh^{-1} \frac{3\sqrt{\Omega_\Lambda}}{2}r,\\
0_\mathbb{X} &=& f^{-1}_\mathbb{X}(0)=t_1.
\ee
Notice that there is some freedom in the choice of $f_\mathbb{Y}$ since one can take a linear function $f_\mathbb{Y}(y)=\alpha y$,
\be
A(t)
&=&
\left(\frac{3}{2}\alpha^{-1/2}\omega_\mathbb{Y}f_\mathbb{X}(t)\right)^{2/3}
,\label{A(t)2}
\ee
which just rescales the parameter $\omega_\mathbb{Y}$. 

Accordingly, (\ref{a(t)1}) solves both (\ref{FR1}) and  (\ref{F}). If $f_\mathbb{Y}(y)=y$,  both Friedman equations are equivalent to
\be
\frac{{\rm D}a(t)}{{\rm D}t}
&=&
\sqrt{\frac{\Omega_M}{a(t)}},\quad t>0_\mathbb{X}.\label{FR1'}.
\ee
The cosmological constant has disappeared from the right-hand-side of the Friedman equation, but is hidden in the non-Newtonian form of the derivative. The fact that $a$ is a map $\mathbb{X}\to \mathbb{Y}$, with the non-Diophantine projective arithmetic defined in $\mathbb{X}=\mathbb{R}$ by (\ref{43}), and the usual Diophantine arithmetic in $\mathbb{Y}=\mathbb{R}$, implies that the metric tensor $g_{ab}$ should be regarded as a map $g_{ab}:\mathbb{X}^4\to \mathbb{R}$.

\section{Arithmetic analogue of dark matter}
\label{Sec 5}

Kolmogorov, Rashevsky and Burgin contemplated non-Diophantine arithmetics of natural numbers. In non-Newtonian calculus we deal with non-Diophantine arithmetics of real numbers. The argument on practical indistinguishability of super-large numbers can be equally well applied to numbers that are very small. Indeed, one can argue that $2+2=4$ is as obvious as $4/2=2$. However, is 
$$
2^{-100^{100^{100}}}/2=2^{-100^{100^{100}}-1}
$$ 
equally obvious? Similarly to machine infinity $N_\infty$ one can speak of machine zero, a nonzero finite number $N_0$ satisfying $N_0/2=N_0$ (e.g. $N_0=1/N_\infty$). In non-Diophantine projective arithemetic zero is the neutral element of addition, $0_\mathbb{X}=f^{-1}_\mathbb{X}(0)$. No matter which bijection $f_\mathbb{X}:\mathbb{X}\to \mathbb{R}$ one takes, one always finds $x\oplus_\mathbb{X} 0_\mathbb{X}=x$, Such a zero can be non-zero in the ordinary Diophantine sense, so is a natural candidate for $N_0$.

Let us consider a simple example. Actually, the example is so simple that it might seem it cannot produce anything interesting: $\mathbb{X}=\mathbb{R}$, $f_\mathbb{X}(x)=x-\epsilon$, 
$f^{-1}_\mathbb{X}(r)=r+\epsilon$. Here $\epsilon$ is an arbitrary real number, for example $\epsilon=N_0$.
The neutral elements of addition and multiplication are $0_\mathbb{X}=f^{-1}_\mathbb{X}(0)=\epsilon$, 
$1_\mathbb{X}=f^{-1}_\mathbb{X}(1)=1+\epsilon=1+0_\mathbb{X}$. 
Analogously, all real numbers will satisfy $r_\mathbb{X}=f^{-1}_\mathbb{X}(r)=r+0_\mathbb{X}$.
To simplify notation let us skip the index $\mathbb{X}$ in the bijection and in the arithmetic operations it generates,
\be
x\oplus x' = f^{-1}\big(f(x)+f(x')\big)=x+x'-\epsilon,\\
x\ominus x' = f^{-1}\big(f(x)-f(x')\big)=x-x'+\epsilon,\\
x\odot x' = f^{-1}\big(f(x)\cdot f(x')\big)=(x-\epsilon)(x'-\epsilon)+\epsilon,\\
x\oslash x' = f^{-1}\big(f(x)/f(x')\big)=(x-\epsilon)/(x'-\epsilon)+\epsilon.
\ee
`Minus $x$' is given by 
\be
\ominus x=0_\mathbb{X}\ominus x
= f^{-1}\big(-f(x)\big)=-x+2\epsilon.
\ee
Let us cross-check:
\be
\ominus x\oplus x=(-x+2\epsilon)+x-\epsilon=\epsilon=0_\mathbb{X}
\ee
as required.  An arbitrary real power of $x$ reads
\be
x^{r_\mathbb{X}}=f^{-1}\big(f(x)^r\big)=(x-\epsilon)^r+\epsilon.
\ee 
It satisfies the usual rules
\be
x^{r_\mathbb{X}}\otimes x^{s_\mathbb{X}} &=&f^{-1}\big(f(x)^rf(x)^s\big)\\
&=& x^{(r+s)_\mathbb{X}}=(x-\epsilon)^{r+s}+\epsilon,
\ee
and
\be
\frac{{\rm D}x^{r_\mathbb{X}}}{{\rm D}x} &=&f^{-1}\big(rf(x)^{r-1}\big)\\
&=&
r_\mathbb{X}\otimes x^{(r-1)_\mathbb{X}}
=r(x-\epsilon)^{r-1}+\epsilon.
\ee
Let us apply the above formulas to a general non-Newtonian Newton-Coulomb `$1/x$' potential
\be
U(x)=x^{(-1)_\mathbb{X}}=x^{\ominus 1_\mathbb{X}}=f^{-1}\big(1/f(x)\big).
\ee
The non-Newtonian force
\be
\ominus
\frac{{\rm D}U(x)}{{\rm D}x}
&=&
f^{-1}\big(1/f(x)^2\big)
=
\frac{1}{(x-\epsilon)^2}+\epsilon\\
&=&
\frac{1}{(x-0_\mathbb{X})^2}+0_\mathbb{X}
\ee
has singularity at $x=0_\mathbb{X}$, and tends to $0_\mathbb{X}$ with $x\to\infty$. 

In our example both Diophantine and non-Diophantine arithmetics, and both Newtonian and non-Newtonian calculi, are defined in the same set $\mathbb{R}$. All the formulas can be thus read in terms of either of them. But since the laws of  physics are formulated here in a non-Diophantine/non-Newtonian way, attempts of interpreting them in the Diophantine/Newtonian formalism will lead to inconsistencies. For example, the attractive gravitational force will asymptotically achieve a constant value $0_\mathbb{X}=\epsilon$ and not just 0. An appropriately formulated centrifugal force on a circular orbit, as well as the linear velocity along the orbit, will also asymptotically tend to $0_\mathbb{X}=\epsilon$, and not just to 0. The mismatch of the two arithmetics will have observable consequences analogous to those of a dark matter.

In the next Section we solve step by step the problem of  velocity on a circular orbit around mass $M$. The solution is valid in any projective arithmetic. One can analogously formulate all of the standard `Newtonian physics' by means of a non-Newtonian calculus. All the formulas will have the usual textbook form. One only will have to replace ${\rm d}x/{\rm d}t$ by ${\rm D}x/{\rm D}t$, $+$ by $\oplus$, 0 by $0_\mathbb{X}$, and so on and so forth. 

What is even more important, one can analogously reformulate any theory which is based on some form of a calculus. General relativity will not be an exception from the rule. Paradoxically, if needed, one could consider mathematically non-Newtonian versions of physically non-Newtonian theories such as MOND or MOG. 

\section{Non-Newtonian velocity on a circular orbit}
\label{App}

Arithmetic works with dimensionless variables, so we need dimensional `fundamental units' (denoted by the Gothic font). For example:
position $(x,y,z)\mathfrak{l}$, velocity $(\dot x,\dot y,\dot z)\mathfrak{v}$, acceleration $(\ddot x,\ddot y,\ddot z)\mathfrak{a}$, time $t\mathfrak{t}$, mass $m\mathfrak{m}$. We do not yet specify which units are truly fundamental. Velocity perhaps satisfies $\mathfrak{v}=c$, but at this stage we leave it arbitrary. The arithmetic is projective in the sense of Burgin, i.e. is defined by means of some one-to-one $f:\mathbb{X}\to \mathbb{R}$. We assume that arithmetics of domains and images of the maps in question are identical (which is not obvious, so this is an assumption about this concrete model). By this it is meant that, for example $\mathbb{X}\ni t\mapsto x(t)\in \mathbb{X}$, so $\dot x(t)={\rm D} x(t)/{\rm D}t\in \mathbb{X}$. Employing our previous notation we write $r_\mathbb{X}=f^{-1}(r)\in \mathbb{X}$ for $r\in \mathbb{R}$. The same concerns the infinity $\infty_\mathbb{X}=f^{-1}(\infty)$. Elements of $\mathbb{X}$ are ordered by $x\le_\mathbb{X} x'$ iff $f(x)\le f(x')$. Minus means $\ominus x=0_\mathbb{X}\ominus x$. In particular, minus infinity is $\ominus\infty_\mathbb{X}=f^{-1}(-\infty)$.  Recalling that an $n$th power of $x\in \mathbb{X}$ is denoted by $x^{n_\mathbb{X}}$ one should similarly denote higher derivatives by
${\rm D}^{n_\mathbb{X}} x(t)/{\rm D}t^{n_\mathbb{X}}$, but I prefer the less redundant, simpler and yet unambiguous form 
${\rm D}^n x(t)/{\rm D}t^n$. The chain rule for derivatives of compositions of functions $A,B:\mathbb{X}\to \mathbb{X}$ reads \cite{Czachor2019}
\be
\frac{{\rm D}A(B(t))}{{\rm D}t}
=
\frac{{\rm D}A\big( B(t)\big)}{{\rm D}B(t)}
\otimes
\frac{{\rm D}B(t)}{{\rm D}t}
.
\ee
The non-Newtonian Hamilton equations
\be
\frac{{\rm D}p_a(t)}{{\rm D}t}
&=&
\ominus \frac{{\rm D}H(q(t),p(t))}{{\rm D}q^a(t)},\\
\frac{{\rm D}q^a(t)}{{\rm D}t}
&=&
\frac{{\rm D}H(q(t),p(t))}{{\rm D}p_a(t)},
\ee
imply the non-Newtonian Poisson bracket
\be
\{A,B\}
&=&
\frac{{\rm D}A}{{\rm D}q^a}\otimes\frac{{\rm D}B}{{\rm D}p_a}
\ominus
\frac{{\rm D}B}{{\rm D}q^a}\otimes\frac{{\rm D}A}{{\rm D}p_a}
\ee
(with non-Newtonian summation convention for $\oplus$). Now, let
\be
H &=&
(p_1^{2_\mathbb{X}}\oplus p_2^{2_\mathbb{X}}\oplus p_3^{2_\mathbb{X}})\oslash (2_\mathbb{X}\otimes m_\mathbb{X})
\nonumber\\
&{}&\pp=
\ominus
G_\mathbb{X}\otimes m_\mathbb{X} \otimes M_\mathbb{X}\oslash r\\
&=&
f^{-1}
\left(
\frac{f(p_1)^2+f(p_2)^2+f(p_3)^2}{2m}
\right.
\nonumber\\
&{}&\pp{f^{-1}(}
\left.
-G\frac{mM}{\sqrt{f(x_1)^2+f(x_2)^2+f(x_3)^2}}
\right),
\ee
where 
$r=(x_1^{2_\mathbb{X}}\oplus x_2^{2_\mathbb{X}}\oplus x_3^{2_\mathbb{X}})^{(1/2)_\mathbb{X}}$,
and all the variables are dimensionless. The non-Newtonian Hamilton equations read explicitly
\be
\frac{{\rm D}p_j}{{\rm D}t}
&=&
f^{-1}\left(\frac{-GmMf(x_j)}{\big(f(x_1)^2+f(x_2)^2+f(x_3)^2\big)^{3/2}}\right),\\
\frac{{\rm D}x_j}{{\rm D}t}
&=&
f^{-1}\left(
\frac{f(p_j)}{m}
\right)
=
v_j.
\ee
The analogue of the diagram (\ref{diagram}),
\be
\begin{array}{rcl}
\mathbb{X}                & \stackrel{x_j,p_j}{\longrightarrow}       & \mathbb{X}               \\
f{\Big\downarrow}   &                                     & {\Big\downarrow}f   \\
\mathbb{R}                & \stackrel{\tilde x_j,\tilde p_j}{\longrightarrow}   & \mathbb{R}
\end{array}\label{diagram pq},
\ee
defines the functions $\tilde x_j$, $\tilde p_j$ such that 
\be
x_j(t) &=& f^{-1}\big(\tilde x_j(f(t))\big),\\
p_j(t) &=& f^{-1}\big(\tilde p_j(f(t))\big),\\
\frac{{\rm D}x_j(t)}{{\rm D}t}
&=&
f^{-1}
\left(
\frac{{\rm d}\tilde x_j\big(f(t)\big)}{{\rm d}f(t)}
\right),\\
\frac{{\rm D}p_j(t)}{{\rm D}t}
&=&
f^{-1}
\left(
\frac{{\rm d}\tilde p_j\big(f(t)\big)}{{\rm d}f(t)}
\right).
\ee
The non-Newtonian Hamilton equations can be therefore rewritten as 
\be
\frac{{\rm d}\tilde p_j\big(f(t)\big)}{{\rm d}f(t)}
&=&
\frac{-GmM \tilde x_j(f(t))}{\big(\tilde x_1(f(t))^2+\dots+\tilde x_3(f(t))^2\big)^{3/2}},\\
\frac{{\rm d}\tilde x_j\big(f(t)\big)}{{\rm d}f(t)}
&=&
\frac{\tilde p_j(f(t))}{m}
=
f\big(v_j(t)\big)=\tilde v_j(f(t)),
\ee
which is the standard Newtonian problem for the {\it intermediate\/} quantities $\tilde x_j$ and $\tilde p_j$. However, once we have found them, we still have to compute the {\it observable\/} quantities $x_j(t)$ and $p_j(t)$. 

The circular orbit $r=R_\mathbb{X}=f^{-1}(R)$ is equivalent to 
\be
R^2 = \tilde x_1(f(t))^2+\tilde x_2(f(t))^2+\tilde x_3(f(t))^2,\\
\frac{{\rm d}^2\tilde x_j\big(f(t)\big)}{{\rm d}f(t)^2}
=
-\frac{GM }{R^3}\tilde x_j(f(t))=
-\omega_R{}^2\tilde x_j(f(t)).
\ee
An orbit located at $x_3=0_\mathbb{X}$ corresponds to the circle $R^2 = \tilde x_1(f(t))^2+\tilde x_2(f(t))^2$ and rotation with angular velocity $\omega_R$, so
\be
\tilde v_1(f(t))^2+\tilde v_2(f(t))^2 =\omega_R{}^2 R^2=GM /R.
\ee
Finally,
\be
v&=&(v_1^{2_\mathbb{X}}\oplus v_2^{2_\mathbb{X}}\oplus v_3^{2_\mathbb{X}})^{(1/2)_\mathbb{X}}
=f^{-1}(\sqrt{GM/R})\\
&=&
f^{-1}(\sqrt{GM/f(R_\mathbb{X})})
=
\big(
(GM)_\mathbb{X}\oslash R_\mathbb{X}\big)^{(1/2)_\mathbb{X}}.\nonumber
\ee
With $R_\mathbb{X}\to\infty_\mathbb{X}$ (i.e. $R\to\infty$) the dimensionless velocity tends to $f^{-1}(0)=0_\mathbb{X}$. The asymptotic dimensional velocity is $0_\mathbb{X}\mathfrak{v}$. This is the general solution, valid for any projective non-Diophantine arithmetic. 

Analogous calculations lead to the dimensionless acceleration
\be
a&=&(a_1^{2_\mathbb{X}}\oplus a_2^{2_\mathbb{X}}\oplus a_3^{2_\mathbb{X}})^{(1/2)_\mathbb{X}}
=f^{-1}(GM/R^2)\label{MOND-f}\\
&=&
f^{-1}\big(GM/f(R_\mathbb{X})^2\big)=v^{2_\mathbb{X}}\oslash R_\mathbb{X}.
\ee
The dimensional one is $a\mathfrak{a}$. It tends asymptotically to $0_\mathbb{X}\mathfrak{a}$.

The non-Newtonian formalism predicts that the limiting velocity and acceleration are mass independent, unless mass influences the form of arithmetic, of course. In any case, the rate of convergence toward $0_\mathbb{X}$ with growing $R$ is mass dependent. 

\section{Comparison with MOND}
\label{Sec MOND}

Formula (\ref{MOND-f}) can be compared with Milgrom's MOND. Assume $f^{-1}(x)\approx x$ for $x\gg x_f$ , 
$f^{-1}(x)\approx \sqrt{a_0 x}$ for $0<\epsilon <x<x_f$, and anything else otherwise. It is not difficult to invent a bijection $f:\mathbb{R}\to \mathbb{R}$ with such properties. Then
\be
a(R) \approx
\left\{
\begin{array}{l}
GM/R^2 \quad \textrm{for $GM/R^2\gg x_f$},\\
\sqrt{GM a_0}/R \quad \textrm{for $\epsilon< GM/R^2< x_f$},\\
f^{-1}(0) \quad \textrm{for $R\to\infty$}.
\end{array}
\right.
\ee
For velocity we get
\be
v(R) \approx
\left\{
\begin{array}{l}
\sqrt{GM/R} \quad \textrm{for $\sqrt{GM/R}\gg x_f$},\\
\sqrt{a_0\sqrt{GM/R}} \quad \textrm{for $\epsilon< \sqrt{GM/R}< x_f$},\\
f^{-1}(0) \quad \textrm{for $R\to\infty$}.
\end{array}
\right.
\ee
Denoting $\sqrt{a_0\sqrt{GM/R}}=v_{\rm M}(M,R)$ we obtain a MOND-type prediction
\be
\frac{v_{\rm M}(M,R)^4}{M}
=
\frac{v_{\rm M}(M',R)^4}{M'}
\ee
for any $M$, $M'$, $R$.

A fundamental principle behind MOND is now reduced to the one that governs the form of $f^{-1}$. For the moment the principle is unknown.

\section{Arithmetic universe}
\label{strange}

\begin{flushright}
{\it Out of nothing I have created a strange new universe\/}

J\'anos Bolyai, from a letter to his father (1823)
\end{flushright}

\medskip

Immanuel Kant  in his {\it Critique of Pure Reason\/} (1781) concluded that the concept of space is unique and given {\it a priori\/}. At the moment of publishing the book Gauss was four years old, while Bolyai, Lobachevski, Riemann, and Einstein were not yet born. One cannot blame Kant for his unawareness of different geometries. After Riemann but before Einstein mathematicians already knew that non-Euclidean geometries were possible. According to some accounts, Gauss even made measurements testing if angles in sufficiently large triangles indeed sum to  $\pi$. Curvilinear systems of coordinates were applied to differential equations  in  Euclidean space much earlier, but the principle of general covariance, implicit in the works of  Lam\'e \cite{Lame}, had the status of a mathematical trick used to simplify calculations. Neither did one know that non-Euclidean character of space is experienced under the name of gravity, nor that different geometries are implied by different distributions of matter.

Identical questions can be posed in the context of arithmetic. Are we still in the arithmetic Kantian era, with arithmetic given 
{\it a priori\/}? Certainly not. After the works of Burgin, Grossman and Katz we are already in a Riemannian era, with formalism at hand but with no true applications in mind. 

\begin{figure}\center
\includegraphics[width=8cm]{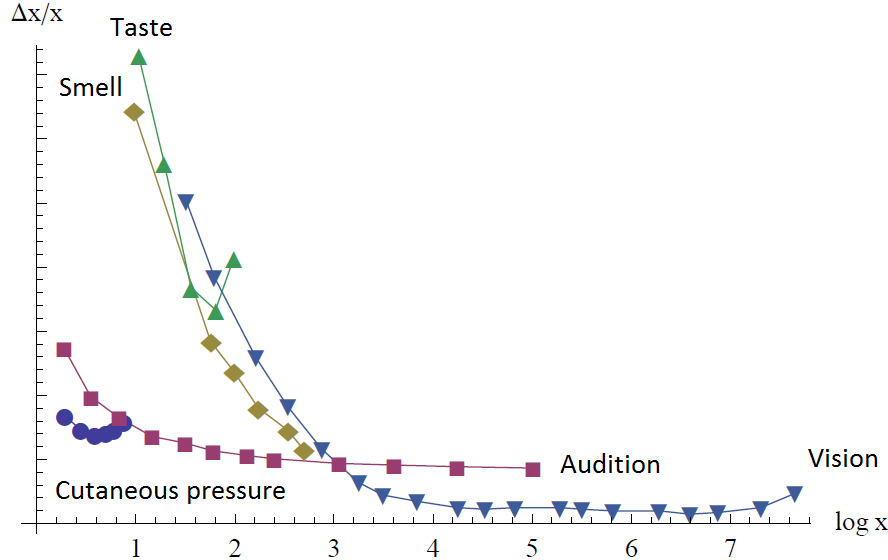}
\includegraphics[width=8cm]{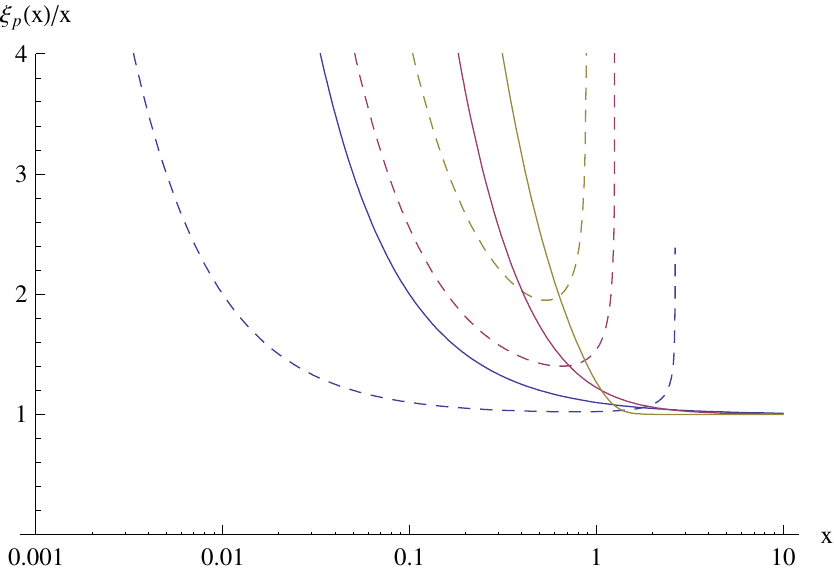}
\caption{Indication of non-Diophantine arithmetic of human sensory system.  Typical Weber-law data versus sensation level of the stimulus (adapted from \cite{LG1963}). $\Delta x/x$ is essentially $\xi_p(x)/x-1$ for some $p$ (division and subtraction are here Diophantine). The traditional Weber law corresponds to the flat parts of the curves. Below, the plots of $\xi_p(x)/x$ for different $f(x)$. Full curves: $f(x)=x$, $f(\delta_p)=0.1$ (leftmost), $f(x)=x^2$, $f(\delta_p)=0.5$ (middle), $f(x)=x^{10}$, $f(\delta_p)=10$ (rightmost). The even powers are restrictions to $x\geq 0$ of the general bijection $f(x)=|x|^q\textrm{sgn}(x)$, $q\in\mathbb{R}$. Dashed curves: $f(x)=\tanh x$ with $f(\delta_p)=0.01$ (lowest), $f(\delta_p)=0.15$ (middle), $f(\delta_p)=0.29$ (upper). The bending-up occurring for the dashed lines is a consequence of bounding $f(x)$ from above. }
\label{Weber_fractions}
\end{figure}

Moreover, most of us is unaware that we in fact {\it do\/} experience non-Diophantine arithmetic in our everyday life. In this respect we behave like those 19th century physicists and mathematicians who experienced gravity but searched in vain for some observable manifestations of non-Euclidean geometry of the Universe. A physical non-Diophantine arithmetic is literally hiding just before our noses:  human and animal sensory systems perform a non-Diophantine subtraction,
\be
\xi_p(x)\ominus x = f^{-1}\big[ f\big(\xi_p(x)\big)-f(x)\big]=\delta_p\label{Fechner}.
\ee
The exact form of the `sensory scale' $f$ is unknown, but experiment shows that a generalized subtraction is at work, making $\delta_p$  independent of $x$ (the Weber law)  \cite{Falmagne}. The so-called sensitivity function $\xi_p(x)$ is, roughly speaking, a perceived value of the input signal $x$, obtained in $p$ percent of measurements. If we know $\delta_p$ and $f$ then
\be
\xi_p(x)
=
f^{-1}\big(f(x)+f(\delta_p)\big)=
x\oplus \delta_p.\label{xi}
\ee
The upper part of Fig.~\ref{Weber_fractions} shows typical Weber-law data for several sensory systems. The lower part illustrates qualitative predictions for various types of non-Diophantine arithmetics \cite{BC}. 
The plateaus are well modeled by $f(x)=a\ln x+b$, a fact explaining why decibels correspond to a logarithmic scale. Non-Diophantine arithmetic {\it is\/} employed by Nature.

Returning to the Universe, it is not easy to accept a scientific paradigm that fills it with huge amounts of unobservable matter, or with pressure $10^{120}$ times smaller than its theoretical estimate. In the non-Newtonian formalism one does not have to change a single scientific law to obtain this type of behavior. Putting it more modestly, even if the arithmetic perspective will not entirely eliminate the need for dark matter or energy, it should at least change theoretical estimates for their parameters.  `Out of nothing' we open new theoretical possibilities. 

The big question remains if there exists a natural law determining the form of arithmetic. The problem was partially addressed by P.~Benioff  \cite{B2002,B2005,B2005b} and, in an explicitly non-Diophantine manner, by J.-C. Falmagne 
\cite{Falmagne2004,Falmagne2015}.  One should also mention the bit-string formalism of H. Pierre Noyes  \cite{Noyes}, and the universal computational rewrite paradigm of P.~Rowlands \cite{Rowlands}.  
All these results, unfortunately, do not seem to bring us any closer to a universal form of $f_\mathbb{X}$, a putative driving force behind our dark Universe.

\section*{Acknowledgments}

I am indebted to Mark Burgin, Mike Grossman, and Peter Carr for encouragement and valuable advice at various stages of this project. I am grateful to Igor Kanatchikov for suggestions how to clarify the presentation.

\section*{Appendix: Fractals as dark universes of zero Lebesgue measure}
\label{Sec 6}

 This section is somewhat orthogonal to the preceding ones, so I decided to shift it to the Appendix. It shows that dark energy may be indeed a real energy that appears `out of nowhere'.  The common element of all these approaches is provided by the non-Newtonian calculus. We will illustrate the idea on one of the simplest fractals: a Cantor dust.

\begin{figure}\center
\includegraphics[width=8cm]{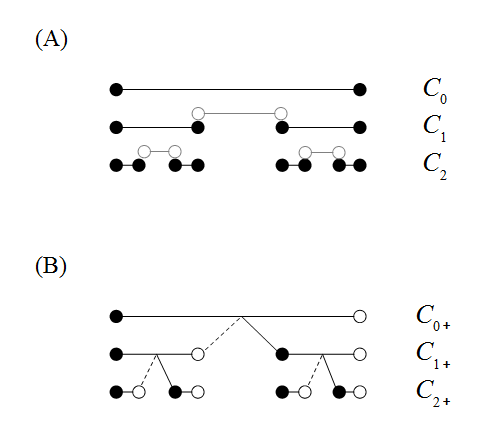}
\caption{Two ways of constructing ternary middle-third Cantor-type fractals. (A) One starts with a closed interval $C_0$. In each step one divides intervals into three segments of equal lengths, and removes interiors of the middle ones. $C_n$ is the set obtained in the $n$-th step. After $n$ steps the number of endpoints equals $2^{n+1}$. The Cantor set is $C=\cap_{n=0}^\infty C_n$. (B) One starts with a right-open interval $C_{0+}$. Each interval is cut in the middle and split, forming two right-open intervals, three times shorter than the split one. $C_{n+}$ is the set obtained in the $n$-th step. After $n$ steps the number of endpoints equals $2^{n}$.The Cantor set is $C_+=\cap_{n=0}^\infty C_{n+}$. Both Cantor sets are self-similar, with the same similarity dimension $\log_3 2$. Clearly, the second procedure is one-to-one so it defines a bijection $f_+:C_+\to C_{0+}$. }
\label{Fig5.1}
\end{figure}

The usual triadic middle-third Cantor set is  constructed by the algorithm from Fig.~\ref{Fig5.1}A. In the first step one removes the interior of the middle one-third of the segment $C_0=[0,1]$. The result is $C_1=[0,1/3]\cup [2/3,1]$. In the second step one performs the same operation on $[0,1/3]$ and $[2/3,1]$, arriving at
$C_2=[0,1/9]\cup[2/9,3/9]\cup[6/9,7/9]\cup[8/9,1]$. And so on,  ad infinitum.  The sets are embeded in one another: $C_0\supset C_1 \supset C_2\supset\dots$.  The Cantor set is the limit $C=\cap_{n=0}^\infty C_n$. The Lebesgue measure $\mu$ of $C_n$ satisfies $\mu(C_{n+1})=\frac{2}{3}\mu(C_n)=\left(\frac{2}{3}\right)^{n+1}\mu(C_0)$, which implies $\mu(C)=\lim_{n\to\infty}\left(\frac{2}{3}\right)^n\mu(C_0)=0$.

One can perform an analogous construction for $C_{0+}=[0,1)$ (Fig.~\ref{Fig5.1}B), but in each step removing a left-closed interval, so that $C_{1+}=[0,1/3)\cup [2/3,1)$, etc. The resulting set $C_+=\cap_{n=0}^\infty C_{n+}$ is self-similar with similarity dimension $d=\log_32$.  `Cantor dusts' $C$ and $C_+$ have the same similarity dimensions. The sets are uncountable.

Fig.~\ref{Fig5.1}B shows why $C_+$ and $[0,1)$ are in a one-to-one relation. Repeating the procedure in any interval $[k,k+1)$, $k=0,\pm 1,\pm 2\dots$, we obtain a subset $C_\mathbb{R}\subset \mathbb{R}$, which is a periodic repetition of $C_+\subset [0,1)$.  $C_\mathbb{R}$ and $\mathbb{R}$ are related by a one-to-one map $f_{C_\mathbb{R}}:C_\mathbb{R}\to\mathbb{R}$. The Lebesgue measure of a countable union of zero-measure sets is zero, so $\mu(C_\mathbb{R})=0$. Readers interested in more explicit details should consult \cite{MC2015,ACK2016a,ACK2016b}

The next step is to consider a Schr\"odinger equation for a wave function $\psi(x)$, $\int_\mathbb{R} |\psi(x)|^2{\rm d}x<\infty$.
$\psi(x)$ is a representative of the {\it equivalence class\/} $|\psi\rangle$ \cite{states}. It can be modified on any zero-Lebesgue-measure set, for example $C_\mathbb{R}$, and yet quantum mechanical measurements will not notice the difference. The fact that the resulting modified wave function may not be differentiable is not a problem. Simply, one defines a derivative $|\psi'\rangle$ of the state $|\psi\rangle$ as the derivative ${\rm d}\psi(x)/{\rm d}x$ of this representative of the equivalence class which is differentiable, and then treats it as a representative of the entire equivalence class $|\psi'\rangle$. Once ${\rm d}\psi(x)/{\rm d}x$ is computed one can arbitrarily modify it at  $x\in C_\mathbb{R}$.

Assume the modification is such that $\psi(x)$ and all its derivatives are set to 0 for $x\in C_\mathbb{R}$. Now we can take an arbitrary $\phi(x)$ which is defined only on  $ C_\mathbb{R}$. We will treat $\phi$  as a completely independent entity, unrelated to $\psi$. For simplicity assume that $\phi(x)\in \mathbb{R}$, while $\mathbb{R}$ is equipped with the ordinary Diophanthine arithmetic. The diagram (\ref{diagram}) then reads
\be
\begin{array}{rcl}
C_\mathbb{R}                & \stackrel{\phi}{\longrightarrow}       & \mathbb{R}               \\
f_{C_\mathbb{R}}{\Big\downarrow}   &                                     & {\Big\downarrow}{\rm id}_\mathbb{R}   \\
\mathbb{R}                & \stackrel{\tilde \phi}{\longrightarrow}   & \mathbb{R}
\end{array}\label{diagram psi}
\ee
Here ${\rm id}_\mathbb{R}(x)=x$ is the identity map. $C_\mathbb{R}$ is equipped with its intrinsic arithmetic,
\be
x\oplus_{C_\mathbb{R}} x' = f^{-1}_{C_\mathbb{R}}\big(f_{C_\mathbb{R}}(x)+f_{C_\mathbb{R}}(x')\big),
\ee
etc. The derivative is just
\be
\frac{{\rm D}\phi(x)}{{\rm D}x}
&=&
\frac{{\rm d}\tilde \phi\big(f_{C_\mathbb{R}}(x)\big)}{{\rm d}f_{C_\mathbb{R}}(x)}.
\label{DA_C}
\ee

An  energy eigenvalue for a quantum system defined on $C_\mathbb{R}$ is given by the Schr\"odinger equation for $\phi:C_\mathbb{R}\to \mathbb{R}$,
\be
-\frac{{\rm D}^2\phi(x)}{{\rm D}x^2}
+
U(x)\phi(x)
=
E\phi(x),\label{dark e}
\ee
or equivalently
\be
-\frac{{\rm d}^2\tilde \phi\big(f_{C_\mathbb{R}}(x)\big)}{{\rm d}f_{C_\mathbb{R}}(x)^2}
+
\tilde U\big(f_{C_\mathbb{R}}(x)\big)\tilde \phi\big(f_{C_\mathbb{R}}(x)\big)
=
E\tilde \phi\big(f_{C_\mathbb{R}}(x)\big).\nonumber\\
\label{dark e1}
\ee
The potential $\tilde U$ is defined by the diagram 
\be
\begin{array}{rcl}
C_\mathbb{R}                & \stackrel{U}{\longrightarrow}       & \mathbb{R}               \\
f_{C_\mathbb{R}}{\Big\downarrow}   &                                     & {\Big\downarrow}{\rm id}_\mathbb{R}   \\
\mathbb{R}                & \stackrel{\tilde U}{\longrightarrow}   & \mathbb{R}
\end{array}\label{diagram U}
\ee
As we can see, we have to solve the ordinary Schr\"odinger equation with the potential $\tilde U$.
If spectrum is discrete the solution is normalized by
\be
\langle\phi|\phi\rangle
=
\int_{C_\mathbb{R}}
|\phi(x)|^2{\rm D}x
=
\int_{-\infty}^{\infty}
|\tilde \phi(r)|^2{\rm d}r=1.
\ee
The eigenvalue $E$ comes from the zero-measure set $C_\mathbb{R}$. The presence of $\phi(x)$ cannot be discovered by standard quantum measurements performed for $\psi(x)$. Still, the energy $E$ contributes to the overall energy of the system. We can say that (\ref{dark e}) plays a role of a dark energy eigenvalue problem.

The fact that (\ref{dark e1}) is effectively the standard Schr\"odinger equation raises the question if a non-Newtonian modification of quantum mechanics can have less trivial observable consequences. The answer is yes, in principle. In order to understand why, let us consider the harmonic oscillator potential $\tilde U(r)=r^2$, and the non-Diophantine arithmetic in $\mathbb{R}$ defined by  $f(x)=x^3$, 
\be
x\oplus x' &=& f^{-1}\big(f(x)+f(x')\big)=\sqrt[3]{x^3+x'^3},\\
x\ominus x' &=& f^{-1}\big(f(x)-f(x')\big)=\sqrt[3]{x^3-x'^3},\\
x\odot x' &=& f^{-1}\big(f(x)\cdot f(x')\big)=x\cdot x',\\
x\oslash x' &=& f^{-1}\big(f(x)/f(x')\big)=x/ x'.
\ee
Multiplication, division, and the neutral elements are unchanged: $f^{-1}(0)=\sqrt[3]{0}=0$, $f^{-1}(1)=\sqrt[3]{1}=1$.
The equation to solve is
\be
-\frac{{\rm d}^2\tilde \psi(r)}{{\rm d}r^2}
+
r^2\tilde \psi(r)
=
E\tilde \psi(r).
\ee
If $E$ is the minimal energy then $\tilde \psi(r)$ is the Gaussian normalized by $\int_{-\infty}^{\infty}|\tilde \psi(r)|^2{\rm d}r=1$. However, even if we assume that $\psi(x)$ is Diophantine-arithmetic-valued, the probability of finding the particle in $[a,b]$ is given by the non-Newtonian integral
\be
\int_a^b |\psi(x)|^2{\rm D}x=\int_{f(a)}^{f(b)} |\tilde \psi(r)|^2{\rm d}r
=\int_{a^3}^{b^3} |\tilde \psi(r)|^2{\rm d}r,
\ee
since the domain of $\psi$ is equipped with the non-Diophantine arithmetic. Probability of finding a particle in $[a,b]$ is thus given by the integral of the Gaussian over $[a^3,b^3]$, and not over $[a,b]$. Such differences  in principle can be measured. Of course, the argument is valid in all non-Newtonian probabilistic theories, not only in quantum mechanics.


\begin{thebibliography}{99}
\bibitem{Ex1}Exercise: check it explicitly by means of (\ref{ar1})-(\ref{ar4}).
\bibitem{states}$\langle f|g\rangle$ is a scalar product if $\langle f|f\rangle=0$ implies that $|f\rangle$ is a zero vector. However, if $\langle f|f\rangle=\int |f(x)|^2{\rm d}x$ then it is easy to give examples of nonzero functions $f$ which nevertheless imply $\int |f(x)|^2{\rm d}x=0$ if ${\rm d}x$ is the Lebesgue measure on $\mathbb R$. So, although $f(x)\neq 0$ for some $x\in \mathbb R$, the corresponding state-vector $|f\rangle$  must be regarded as {\it the\/} zero vector. In order to make it formal, we introduce the equivalence relation: $f\sim g$ if and only if $f(x)-g(x)$ is nonzero on at most a set of Lebesgue measure zero. The ket $|f\rangle$ is the equivalence class of functions defined by the relation. In consequence, if we modify the values $f(x)$ on a set of zero Lebesgue measure, the state $|f\rangle$ remains unchanged, and thus no quantum measurement can detect the modification. Sets of zero Lebesgue measure can have the same cardinality as the continuum 
$\mathbb R$.
\end{thebibliography}
\end{document}